\documentclass[twocolumn,prl,aps,reprint,nofootinbib,superscriptaddress]{revtex4-2}

\usepackage[utf8]{inputenc}
\usepackage{epsfig}
\usepackage{booktabs}
\usepackage{footnote}
\usepackage{ctable}
\usepackage{comment}
\usepackage{caption,subcaption}
\usepackage[intlimits]{amsmath}
\usepackage{mathtools}
\usepackage{dcolumn}
\usepackage{bm}
\newcommand {\ra} [1] {\renewcommand{\arraystretch}{#1}}
\newcommand {\beq} {\begin{eqnarray}}
\newcommand {\eeq} {\end{eqnarray}}
\newcommand {\eeqn} [1] {\label{#1} \end{eqnarray}}

\usepackage{multirow}
\usepackage{graphicx}
\usepackage{txfonts}
\newcommand\T{\rule{0pt}{2.6ex}}       
\newcommand\B{\rule[-1.2ex]{0pt}{0pt}} 

\begin{document}

\title{Quenching of the $\pi0p_{3/2}-\pi0p_{1/2}$  spin-orbit splitting in $^{20}$O and the effect of the tensor force}

\author{J. Lois-Fuentes}
\altaffiliation[ ]{Corresponding author}
\affiliation{IGFAE and Dpt. de F\'{i}sica de Part\'{i}culas, Univ. of Santiago de Compostela, E-15758, Santiago de Compostela, Spain}
\author{B. Fern\'{a}ndez-Dom\'{i}nguez}
\altaffiliation[]{Corresponding author}
\affiliation{IGFAE and Dpt. de F\'{i}sica de Part\'{i}culas, Univ. of Santiago de Compostela, E-15758, Santiago de Compostela, Spain}
\author{T. Roger}
\affiliation{GANIL, CEA/DRF-CNRS/IN2P3, Bd. Henri Becquerel, BP 55027, F-14076 Caen, France}
\author{F. Delaunay}
\affiliation{LPC Caen UMR6534, Universit\'e de Caen Normandie, ENSICAEN, CNRS/IN2P3, F-14000 Caen, France}
\author{M. Lozano-Gonz\'{a}lez}
\affiliation{IGFAE and Dpt. de F\'{i}sica de Part\'{i}culas, Univ. of Santiago de Compostela, E-15758, Santiago de Compostela, Spain}
\author{O. Sorlin}
\affiliation{GANIL, CEA/DRF-CNRS/IN2P3, Bd. Henri Becquerel, BP 55027, F-14076 Caen, France}
\author{T. Otsuka}
\affiliation{CNS, University of Tokyo, 7-3-1 Hongo, Bunkyo-ku, Tokyo, Japan}
\author{T. Suzuki}
\altaffiliation[Present address: ]{Department of Physics, College of Humanities and Sciences, Nihon University, Sakurajosui 3, Setagaya-ku, Tokyo 156-8550, Japan and School of Physics, Beihang University, 37 Xueyuan Road, Haidian District, Beijing 100191, People’s Republic of China}
\affiliation{NAT Research Center, NAT Corporation, 3129-45 Hibara, Muramatsu, Tokai, Naka, 319-1112, Ibaraki, Japan}
\author{N.L. Achouri}
\affiliation{LPC Caen UMR6534, Universit\'e de Caen Normandie, ENSICAEN, CNRS/IN2P3, F-14000 Caen, France}
\author{M. Caama\~{n}o}
\affiliation{IGFAE and Dpt. de F\'{i}sica de Part\'{i}culas, Univ. of Santiago de Compostela, E-15758, Santiago de Compostela, Spain}
\author{C. Cabo}
\affiliation{IGFAE and Dpt. de F\'{i}sica de Part\'{i}culas, Univ. of Santiago de Compostela, E-15758, Santiago de Compostela, Spain}
\author{L. Caceres}
\affiliation{GANIL, CEA/DRF-CNRS/IN2P3, Bd. Henri Becquerel, BP 55027, F-14076 Caen, France}
\author{A. Candiello}
\affiliation{KU Leuven, Instituut voor Kern- en Stralingsfysica, Celestijnenlaan 200d, 3001 Leuven, Belgium}
\author{A. Cassisa}
\affiliation{Nuclear Physics Institute of the Czech Academy of Sciences,  Husinec-\v{R}e\v{z} CZ-25068 No. 130, Czech Republic}

\affiliation{ Charles University, Faculty of Mathematics and Physics Ke Karlovu 3, 121 16 Praha 2, Czech Republic}

\author{A. Ceulemans}
\affiliation{KU Leuven, Instituut voor Kern- en Stralingsfysica, Celestijnenlaan 200d, 3001 Leuven, Belgium}
\author{F. Cresto}
\affiliation{Laboratoire de Physique des deux Infinis Bordeaux (LP2IB), UMR 5797, CNRS/IN2P3, Universit\'e de Bordeaux, Chemin du Solarium, F-33170 Gradignan, France}
\author{Q. Delignac}
\affiliation{Laboratoire de Physique des deux Infinis Bordeaux (LP2IB), UMR 5797, CNRS/IN2P3, Universit\'e de Bordeaux, Chemin du Solarium, F-33170 Gradignan, France}
\author{J.A. Due\~nas}
\affiliation{ Departamento de Ingenier\'{i}a El\'ectrica y Centro de Estudios Avanzados en F\'{i}sica, Matem\'aticas y Computaci\'on, Universidad de Huelva, 21071 Huelva, Spain.   }
\author{D. Fern\'{a}ndez-Fern\'{a}ndez}
\affiliation{IGFAE and Dpt. de F\'{i}sica de Part\'{i}culas, Univ. of Santiago de Compostela, E-15758, Santiago de Compostela, Spain}
\author{S. Fracassetti}
\affiliation{KU Leuven, Instituut voor Kern- en Stralingsfysica, Celestijnenlaan 200d, 3001 Leuven, Belgium}
\author{J. Giovinazzo}
\affiliation{Laboratoire de Physique des deux Infinis Bordeaux (LP2IB), UMR 5797, CNRS/IN2P3, Universit\'e de Bordeaux, Chemin du Solarium, F-33170 Gradignan, France}
\author{S. Gr\'evy}
\affiliation{Laboratoire de Physique des deux Infinis Bordeaux (LP2IB), UMR 5797, CNRS/IN2P3, Universit\'e de Bordeaux, Chemin du Solarium, F-33170 Gradignan, France}
\author{G.F. Grinyer}
\affiliation{Department of Physics, University of Regina, Regina, SK S4S 0A2, Canada}
\author{V. Guimar\~{a}es}
\affiliation{Departamento de F\'{i}sica Nuclear, Instituto de Fı\'{i}sica da Universidade de S\~{a}o Paulo, CP 66318, 05315-970 São Paulo SP, Brazil}
\author{O. Kamalou}
\affiliation{GANIL, CEA/DRF-CNRS/IN2P3, Bd. Henri Becquerel, BP 55027, F-14076 Caen, France}
\author{T. Kurtukian-Nieto}
\affiliation{Laboratoire de Physique des deux Infinis Bordeaux (LP2IB), UMR 5797, CNRS/IN2P3, Universit\'e de Bordeaux, Chemin du Solarium, F-33170 Gradignan, France}
\author{M.B. Latif}
\affiliation{KU Leuven, Instituut voor Kern- en Stralingsfysica, Celestijnenlaan 200d, 3001 Leuven, Belgium}
\author{B. Mauss}
\affiliation{GANIL, CEA/DRF-CNRS/IN2P3, Bd. Henri Becquerel, BP 55027, F-14076 Caen, France}
\author{C. Nicolle}
\affiliation{GANIL, CEA/DRF-CNRS/IN2P3, Bd. Henri Becquerel, BP 55027, F-14076 Caen, France}
\author{A. Ortega-Moral}
\affiliation{Laboratoire de Physique des deux Infinis Bordeaux (LP2IB), UMR 5797, CNRS/IN2P3, Universit\'e de Bordeaux, Chemin du Solarium, F-33170 Gradignan, France}
\author{J. Pancin}
\affiliation{GANIL, CEA/DRF-CNRS/IN2P3, Bd. Henri Becquerel, BP 55027, F-14076 Caen, France}
\author{J. Piot}
\affiliation{GANIL, CEA/DRF-CNRS/IN2P3, Bd. Henri Becquerel, BP 55027, F-14076 Caen, France}
\author{O. Poleshchuk}
\affiliation{KU Leuven, Instituut voor Kern- en Stralingsfysica, Celestijnenlaan 200d, 3001 Leuven, Belgium}
\author{R. Raabe}
\affiliation{KU Leuven, Instituut voor Kern- en Stralingsfysica, Celestijnenlaan 200d, 3001 Leuven, Belgium}
\author{D. Ramos}
\affiliation{GANIL, CEA/DRF-CNRS/IN2P3, Bd. Henri Becquerel, BP 55027, F-14076 Caen, France}
\author{D. Regueira-Castro}
\affiliation{IGFAE and Dpt. de F\'{i}sica de Part\'{i}culas, Univ. of Santiago de Compostela, E-15758, Santiago de Compostela, Spain}
\author{A.M.  S\'{a}nchez-Ben\'{i}tez}
\affiliation{Departamento de Ciencias Integradas y Centro de Estudios Avanzados en F\'{i}sica, Matem\'aticas y Computaci\'on, Universidad de Huelva, 21071 Huelva, Spain}
\author{J.C. Thomas}
\affiliation{GANIL, CEA/DRF-CNRS/IN2P3, Bd. Henri Becquerel, BP 55027, F-14076 Caen, France}
\author{M. Vandebrouck}
\affiliation{IRFU, CEA, Universit\'{e} Paris-Saclay, F-91191 Gif-sur-Yvette, France}
\author{J.C. Zamora}
\affiliation{Departamento de F\'{i}sica Nuclear, Instituto de Fı\'{i}sica da Universidade de S\~{a}o Paulo, CP 66318, 05315-970 São Paulo SP, Brazil}

\begin{abstract}

We present the first direct measurement of the Z=6 shell gap in the neutron-rich $^{20}$O nucleus. The one-proton removal transfer reaction $^{2}$H($^{20}$O,$^{3}$He)$^{19}$N has been studied using the ACTAR TPC setup at GANIL.  
The use of ACTAR TPC enabled the measurement of low-cross section proton-removal  reactions while preserving resolution. Eight $p$-hole states with $\ell$=1 were identified in $^{19}$N  accounting for total strengths of 86\% and 72\% of the $0p_{3/2}$ and $0p_{1/2}$ single-particle orbitals, respectively. The energies and spectroscopic factors of the measured states allowed to determine the proton spin-orbit splitting $\pi0p_{3/2}-\pi0p_{1/2}$  in $^{20}$O.  The Z=6 shell gap  has been established to be 5.30(14) MeV. These findings indicate a reduction of the Z=6 shell gap while adding neutrons to the $sd$-valence orbitals, consistent with the effects of the tensor force predicted by state-of-the-art shell model interaction SFO-tls while at variance with the emergence of a large Z=6 gap observed in other studies.

\end{abstract}

\maketitle

\textit{Introduction-}In quantum mechanics, the coupling between a particle's velocity and its intrinsic spin gives rise to the spin-orbit effect (SO), a mechanism underlying numerous fundamental phenomena in mesoscopic systems \cite{Gal13,Uhlen26,Zut04}. In nuclear physics, a strong SO interaction splits energy levels that share the same orbital angular momentum ($\ell$) but differ in spin orientation ($s$), thereby shaping the nuclear shell structure and producing the well-known sequence of magic numbers \cite{MayerIII, MayerIV}. 
In a simple approach, the SO force is often represented as a one-body potential proportional to the gradient of the nuclear density, highlighting its surface-dominated character. Although introduced into nuclear theory more than seventy years ago, its microscopic origin  \textendash~ ultimately linked to the quark structure of nucleons \textendash ~remains incomplete, and its strength must still be determined empirically.

Since its initial study in stable nuclei, spin-orbit (SO) splitting has been found to be modified in nuclei near the drip lines \cite{Schiffer,Gaudefroy,Burgunder,Orlandi,Chen-exp}. Relativistic mean-field (RMF) calculations predict that, in exotic neutron-rich nuclei, an increase in surface diffuseness leads to a reduction of the splitting between  SO partners \cite{Dobac94,Lalazi98}. More recently, attention has focused on the effect of a reduced proton density in the nuclear interior of the candidate "bubble" nuclei $^{34}$Si \cite{Mutschler}, which may further weaken the SO interaction \cite{Todd04,Quelle09}. A recent study of the neutron 1$p$ SO-splitting near $^{34}$Si carried out at GANIL and Argonne laboratories, led to two different interpretations. The first attributes the reduction of the SO splitting to a central depletion of the proton density, and therefore to a modification of the SO strength \cite{GrassoSO,KarakatsanisSO,SorlinSO}, while the latter explains the experimental results by invoking finite binding effects arising from the extended radial wave functions. \cite{HamamotoSO,Kay,Chen-so}.

Furthermore, other mechanisms at play are shown to strongly alter the single-particle spectra in exotic nuclei. The proton-neutron tensor interaction has been found to be responsible for the disappearance of traditional magic numbers and the emergence of new ones (see \cite{SorlinIII} and references therein). In addition,  continuum effects are expected to play an important role \cite{Dobac96,Meng99} as experimental studies extend to nuclei away from stability.  To obtain a comprehensive understanding of these modifications, new experimental data are crucial.

In stable nuclei,  the first SO magic number that appears as a consequence of SO-splitting is Z,N= 6. This SO-splitting was already considered by G. Mayer to be rather weak compared to traditional SO magic numbers. However, a recent publication \cite{Ong} points to the appearance of a possible proton-subshell closure at Z=6 in n-rich carbon isotopes. Based on charge radii, electromagnetic transition rates and atomic masses, the authors argue that  Z=6, arising from the proton $0p$ SO-splitting, is comparable to other traditional spin-orbit-driven magic numbers, such as 28,  in stable nuclei. 

However, a reduction of the SO splitting in neutron-rich carbon isotopes, attributed to the tensor force, had previously been proposed by A.O. Macchiavelli et al., \cite{Macchiavelli} to account for the enhanced collectivity observed in $^{20}$C, similar to the 1$p_{1/2}$-1$p_{3/2}$ reduction reported in neutron rich Sr-Zr isotopes  \cite{Federman}. This interpretation is supported by (p,2p) measurements at GSI performed by I. Syndikus et al. \cite{Syndikus}, which quantified the increase in the proton component of the 2$^{+}$ state in n-rich carbon isotopes. However, none of these studies directly measured the proton single-particle energies as neutrons were added to the $sd$-orbitals, which is a key piece of information. Here, we present the first measurement of the 0$p$ excitation energies and corresponding spectroscopic factors in $^{20}$O where fewer correlations are expected than in $^{21}$N.  The experiment demonstrates the capabilities of next generation of active targets in single-nucleon transfer reactions.  The use of ACTAR TPC was crucial, as it enabled compensation for the loss of luminosity inherent to the use of radioactive beams while maintaining high resolution, highlighting its unique advantage for such transfer-reaction studies.

\textit{Experiment-} A radioactive $^{20}$O beam at 35A MeV was produced at GANIL by fragmentation of a 55A MeV $^{22}$Ne primary beam at the object position of the LISE spectrometer \cite{LISE}. After separation in LISE, the secondary beam was delivered to the active-target time-projection-chamber ACTAR TPC \cite{Roger-Demo, Mauss-ACTAR, Jerome-GET}, operated in active-target mode \cite{Mauss-ACTAR}. The $^{20}$O intensity at the target position was about $2.5 \times 10^{4}$ pps, with a purity close to 100\%. The ACTAR TPC allows, by measuring in 3D the vertex of the reaction, the use of relatively thick gas targets without loss of resolution. For this experiment ACTAR TPC was filled in with a mixture of D$_{2}$(90\%) and iC$_{4}$H$_{10}$(10\%) gases at a pressure of 952 mbar at 20$^{\circ}$ C. The gas temperature was monitored continuously during the experiment. The fluctuations were found negligible ($\textless$1\%),  with little effect on the density or the number of atoms in the active volume. With this mixture, the use of the active target ACTAR TPC represents approximately a factor 10 in the number of deuterium atoms compared to standard solid targets used in the past to perform similar (d,$^3$He) measurements \cite{Matta_d3He}.

\begin{figure}[h]
\includegraphics[scale=0.23]{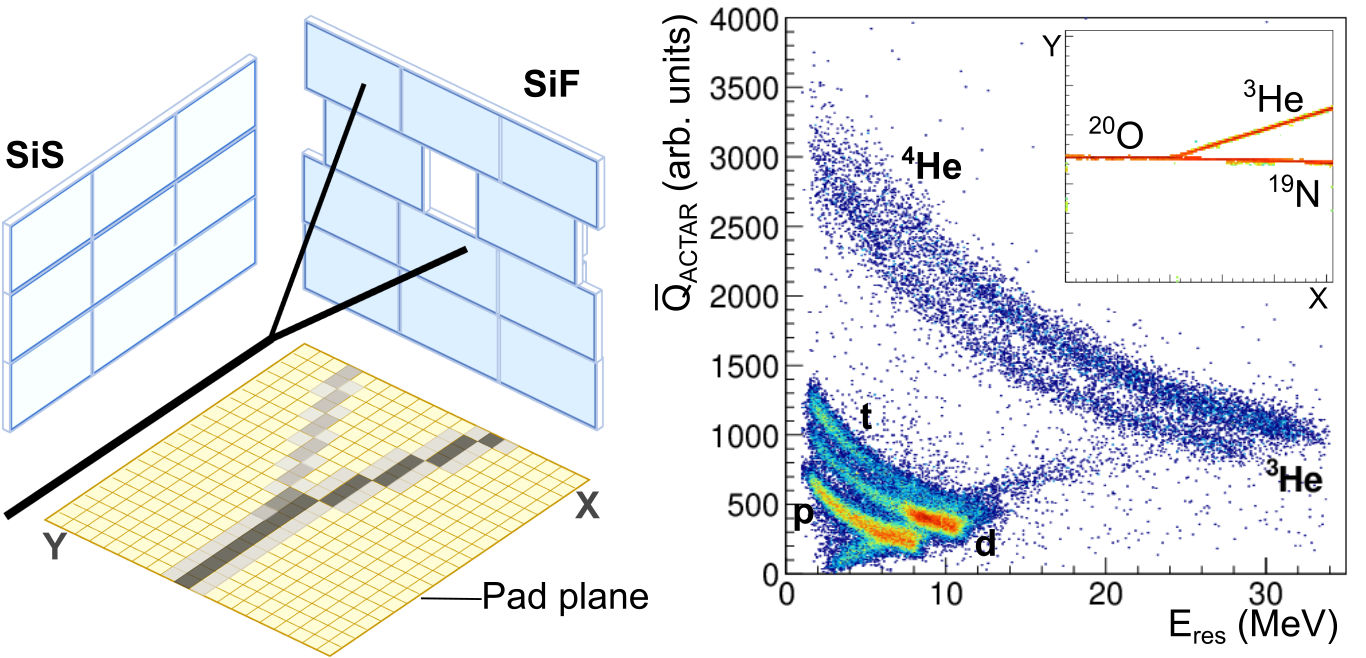}
\caption{\label{PID} (Color online) Left: Scheme of the setup showing the pad plane and both Si walls. Right: The PID is constructed using the energy loss in the gas within ACTAR TPC ($\overline{Q}_{\text{ACTAR}}$) and the residual energy in the first silicon layer (E$_{res}$). The inset shows an example of the projection  of a two-body event in the XY plane of ACTAR TPC.}
\end{figure}

The ACTAR TPC setup is shown in the left panel of Fig.~\ref{PID}. Two silicon detector walls (SiF and SiS) were installed about 10 cm from the field cage, covering forward and transverse laboratory angles to detect light charged particles. Each wall consisted of Si PIN detectors with an active area of $8 \times 5~\mathrm{cm^{2}}$ and a thickness of $500~\mu\mathrm{m}$, providing overall angular coverage from $5^{\circ}$ to $120^{\circ}$\footnote{Assuming the reaction occurs at the center of the ACTAR TPC active volume.}. The emission angles of these particles were obtained directly from the three-dimensional tracks reconstructed within ACTAR TPC, achieving an angular resolution of $\Delta\theta_{\mathrm{lab}} = 0.4^{\circ}$. A multi-wire proportional counter located at the entrance of ACTAR TPC was used to monitor the incoming beam intensity for normalization, which was cross-checked with the elastic-scattering yield measured simultaneously in the side wall (SiS).

The $^{3}$He ions were identified using the $\Delta E$–$E$ technique, with $\Delta E$ derived from the energy deposited in the active volume ($\overline{Q}_{\text{ACTAR}}$) and $E$ from the front silicon detectors (SiF). The right panel of Fig.~\ref{PID} shows the particle identification (PID) of the different light particles. The inset, presents the 2D projection of a typical transfer reaction event, where the beam, recoil and ejectile are clearly distinguishable.

\begin{figure}[h]
\includegraphics[scale=0.35]{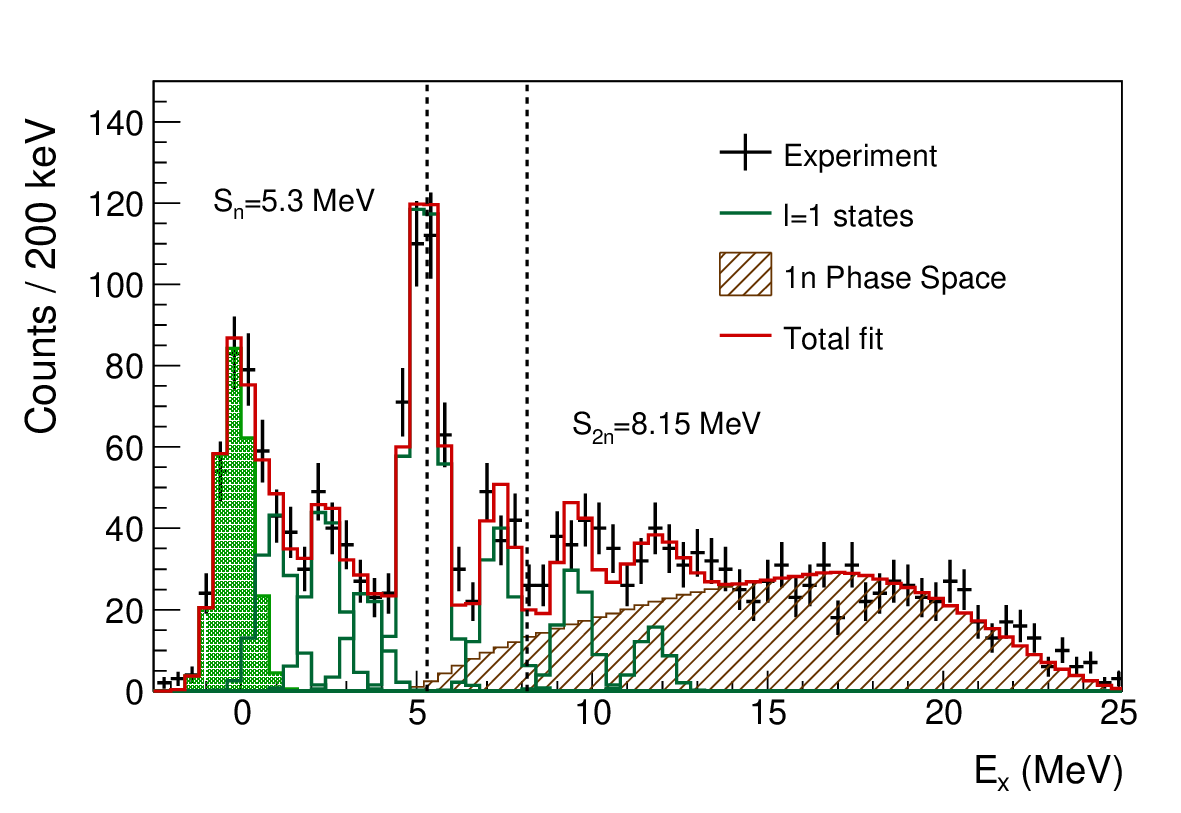}
\caption{\label{nrj_ex} (Color online) Excitation energy spectrum of $^{19}$N reconstructed from the energy and angle of the  $^{3}$He particles.}
\end{figure}

\textit{Results-}The excitation energy  spectrum of $^{19}$N  was extracted from the scattering angles and the kinetic energies of $^{3}$He using the missing-mass technique in the center-of-mass angular range 5$^{\circ}$-22$^{\circ}$. The scattering angle was deduced from a 3D fit to the tracks of the events identified as binary reactions using the multiple-step algorithm developed in \cite{JLois-PhD} (inset of Fig.~\ref{PID} b). By combining the residual energy measured in the silicon detectors with the track length and the energy-loss data obtained from SRIM, we can reconstruct the energy of light particles at the reaction vertex. The final excitation energy (see Fig. \ref{nrj_ex}) shows a typical resolution of 1.06 MeV (FWHM). This value is equivalent to that obtained with a solid target 10 times thinner (1mg/cm$^{2}$).  Bound states were fit with gaussians while unbound states were modelled with Breit-Wigner functions convolved with the experimental resolution, which dominated in all cases.  Additionally, contribution from a three-body final states ($^{18}$N+n+$^{3}$He) was included in the fit. The phase space for the final ($^{17}$N+2n+$^{3}$He) is also included but the fit yields a negligible amplitude. 

Figure \ref{nrj_ex} shows eight clear structures, four states below the neutron separation energy (S$_{n}$=5.328  MeV \cite{AME}) at E$_{x}=0.00(5),1.21(12), 2.58(9)$ and $3.85(15) $ MeV and four resonances above, at E$_{x}=5.30(3),7.29(9)$, $9.39(16)$ and $11.96(26)$ MeV. Except for the ground (1/2$^{-}$) and the first excited state (3/2$^{-}$) reported at 1.143 MeV, the remaining states have not been observed elsewhere \cite{WiltonMultinucleon19N, ENSDF}.

\begin{figure}
\includegraphics[scale=0.35]{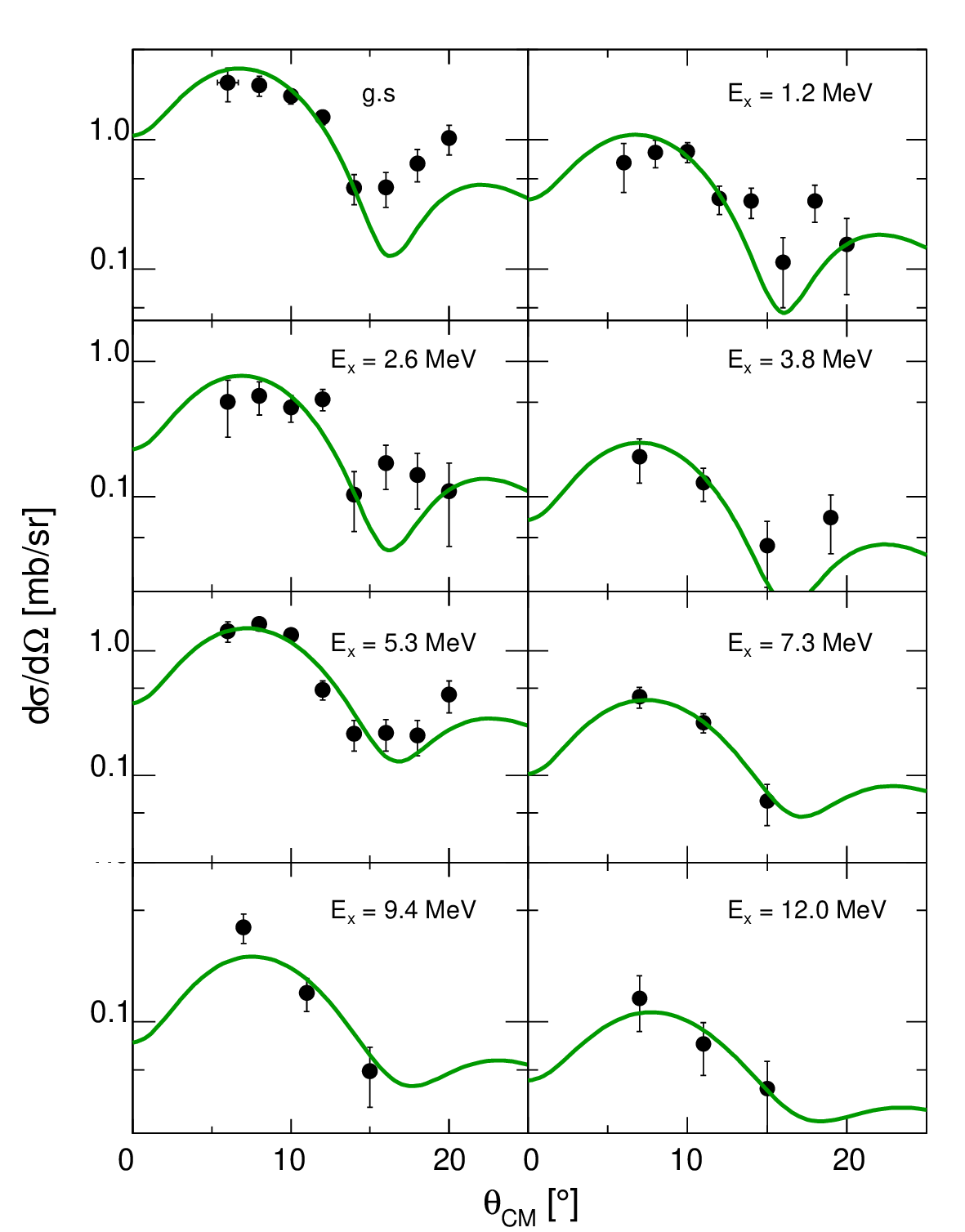}
\caption{\label{ang_dist} Angular distributions for the measured states compared to $\ell$ = 1 (green) DWBA calculations. The uncertainties in the angular distribution are only statistical. With the exception of the final three states, which are doubled, each angular bin corresponds to two degrees. }
\end{figure}

Figure \ref{ang_dist} presents the measured angular distributions for the observed states, compared with zero-range Distorted Wave Born Approximation (DWBA) calculations that incorporate finite-range corrections via the local-energy approximation, computed with the TWOFNR code \cite{TWOFNR}.  For the $d+^{20}$O entrance channel, the optical potential was obtained from the global parameterization of \cite{Daehnick}.  For the  exit channel, the optical potential was taken from Pang \cite{Pang}. The normalization of the $\langle  ^{3}$He $|$  $d +p \rangle$  vertex used a D$_{0}$ constant of -163.8 MeV fm$^{3/2}$ and a spectroscopic factor of 1.31, consistent with the overlap from GFMC calculations \cite{Brida}

The overlap function $\langle ^{20}$O $|$  $^{19}$N $\rangle$ was represented by single particle wave functions obtained in a Woods-Saxon potential of reduced radius r$_{0}$=1.34 fm, diffuseness a=0.65 fm and with the depth adjusted to reproduce the effective proton separation energy. Following Ref. \cite{Hai}, we fix the radius parameter r$_{0}$=1.34  to the global mean of the $r_{0}$ distributions for valence protons obtained from 122 Skyrme interactions. This choice provides consistent reduction factors that reconcile spectroscopic factors (SFs) derived from transfer reactions with those from (e,e$^\prime$p). The SFs reported here are already renormalized by the factor extracted for the ground state of $^{16}$O from (e,e$^\prime$p), R$_s$ = 0.61 \cite{Leuschner, Ryckebusch}.

The shape of the angular distribution provides information on the orbital angular momentum of the removed proton ( $\ell$-value) while the spectroscopic factor (C$^{2}$S) is deduced from the normalisation of the experimental cross section to the calculated one.

All states shown in Fig. \ref{ang_dist} are nicely reproduced by the DWBA calculations assuming a $\ell=1$ proton removal.  This corresponds to the removal of valence protons from the $0p$-shell, as $^{20}$O  is expected to have a robust Z=8 shell closure.  
 The ground state is associated with the removal of a proton from the $0p_{1/2}$ orbital, and therefore, the spin and parity is inferred to be 1/2$^{-}$ in line with earlier studies \cite{Kameda19N, ENSDF}. The corresponding spectroscopic factor C$^{2}$S is 1.66(11) and represents 86\% of the full strength available for the $0p_{1/2}$ orbital. The remaining low-lying states exhibit large C$^{2}$S that would exceed the single-particle occupancy for the $0p_{1/2}$ orbital if a J$^{\pi}$=1/2$^{-}$ assignment were assumed. Consequently, all excited states are assigned J$^{\pi}$=3/2$^{-}$, consistent with a proton removal from the $0p_{3/2}$ orbital. While the single-particle $0p_{1/2}$  strength appears to be essentially concentrated in the ground state, the observed strength of the $0p_{3/2}$ orbital is spread across the excited states and sums up to 72\%.  The relative  $0p_{3/2}$ strength compared to $0p_{1/2}$ is found to be 1.72. Similarly to the one obtained for  $^{16}$O \cite{Mairle16O,Bechtold} of 1.55 and  $^{18}$O \cite{Mairle18O} of 1.62. 
Given the large single-particle strength carried by the states measured in this work, we deduced the effective single-particle energies (ESPE) \cite{Baranger}  of the proton $0p$ orbitals:  $\varepsilon_{\pi0p_{1/2}}$=-19.34(10) MeV and $\varepsilon_{\pi0p_{3/2}}$=-24.64(10) MeV. The spin-orbit splitting is found to be  $\Delta$=$\varepsilon_{\pi0p_{1/2}}-\varepsilon_{\pi0p_{3/2}}$=5.30(14)~MeV in $^{20}$O.

\begin{figure}
\includegraphics[scale=0.40]{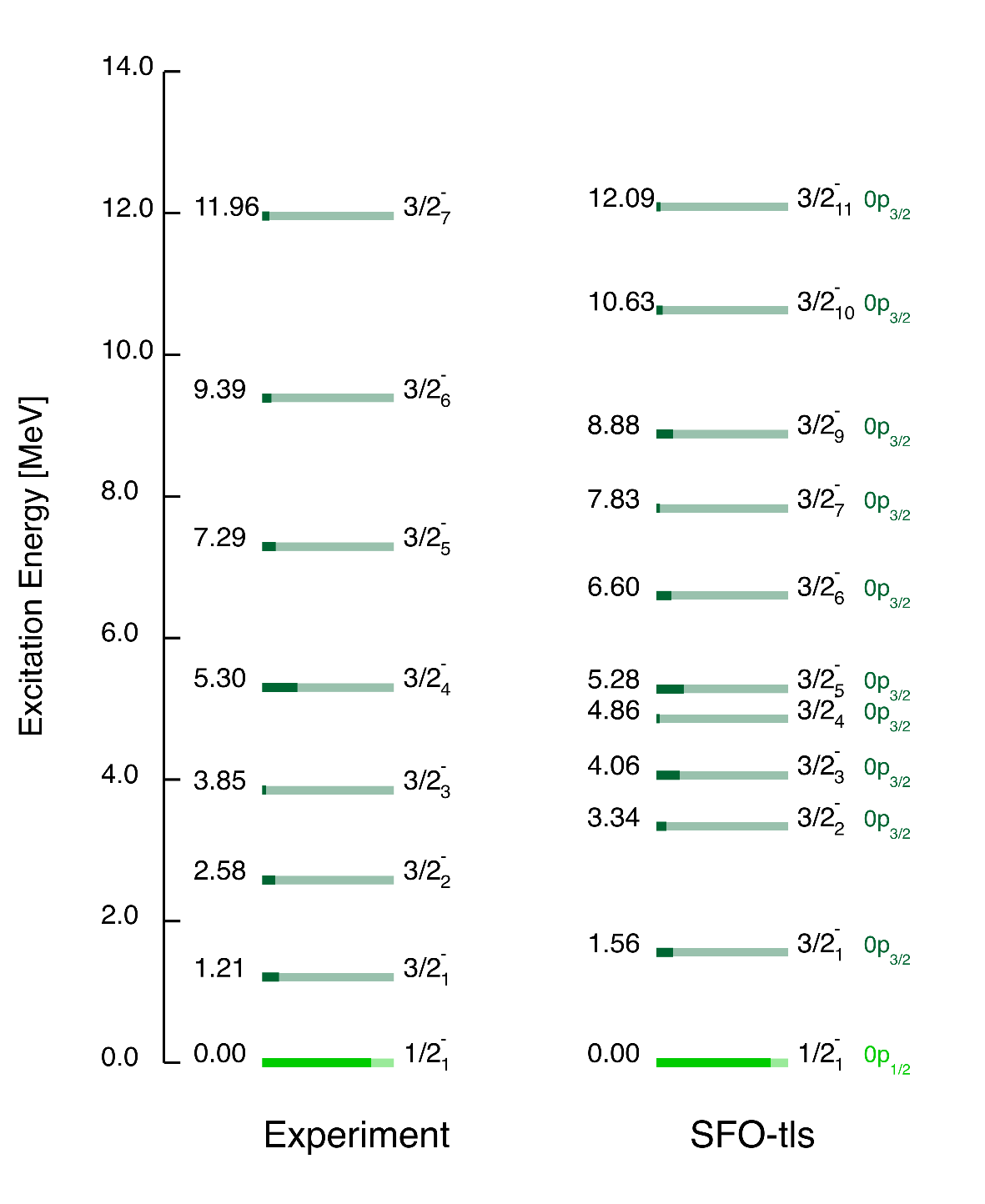}
\caption{\label{ex_nrj_th} Experimental excitation energy of the states measured  in $^{19}$N compared to theoretical calculations with the SFO-tls interaction.  Spins and parities are indicated on the right of the horizontal bars. The filled part of the bar represents the spectroscopic factor C$^{2}$S.  For clarity,  only states with theoretical C$^{2}$S$>$ 0.06 are displayed. }
\end{figure}

\textit{Discussion-} The results were compared to state-of-the-art shell model interactions in the $p-sd$ configuration space using the SFO-tls interaction \cite{TSuzukiI,TSuzukiII, TOtsuka-spe, VMU} (see Fig. \ref{nrj_ex} and Table \ref{tab_sf_19n}). The overall single-proton strength observed experimentally is well reproduced by the SFO-tls interaction. The $0p_{1/2}$ strength is in fairly good agreement with the theory. The experimental distribution of the $0p_{3/2}$  strength is shown to be rather regular Fig. \ref{ex_nrj_th}. However,  the states associated with the $0p_{3/2}$  strength appear more compressed in the shell model calculations. Overall, the summed strength up to the experimentally measured excitation energy is in good agreement with that obtained using the SFO-tls interaction.

\begin{figure}
\includegraphics[scale=0.35]{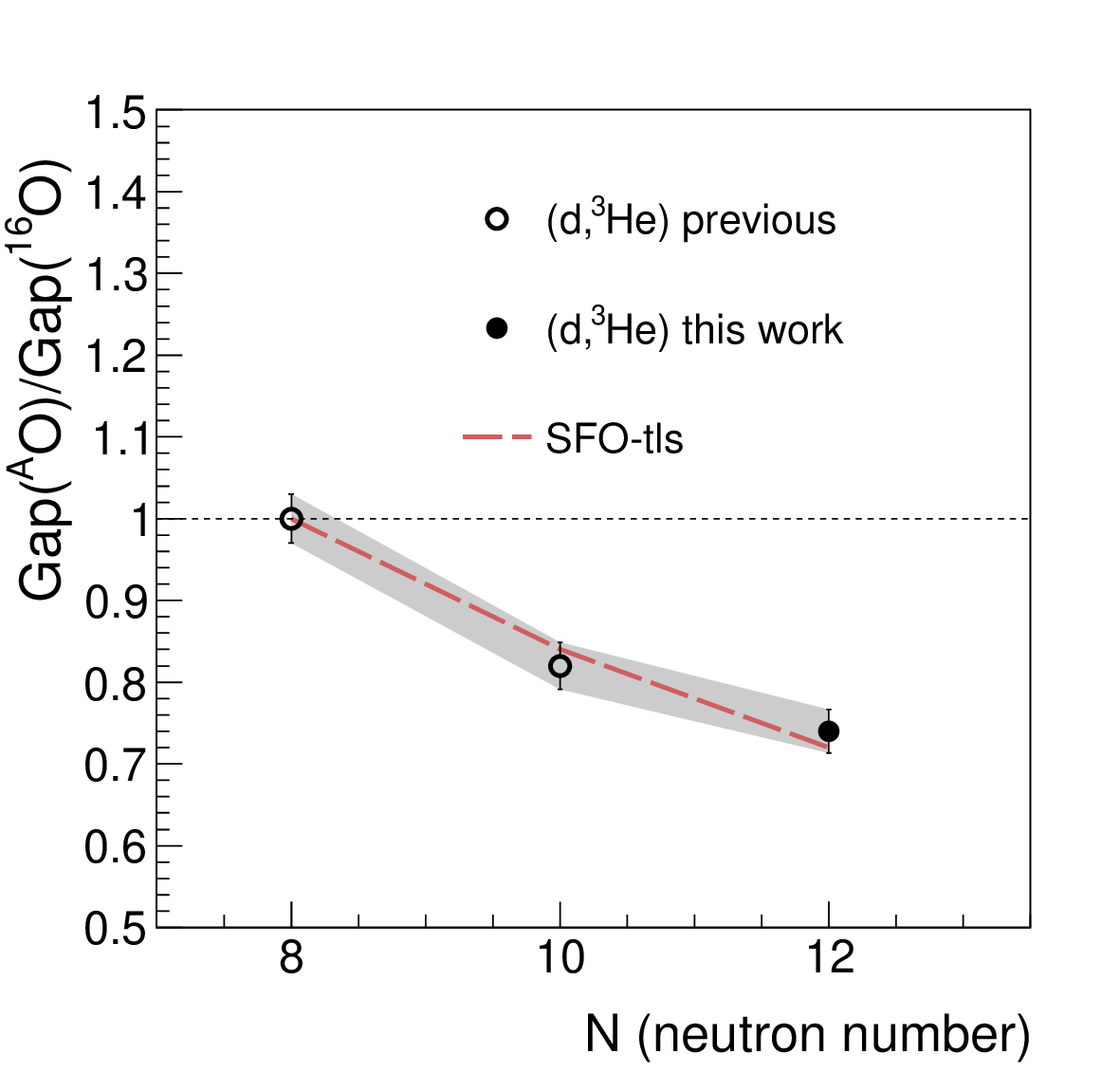}
\caption{\label{espe}  Evolution of the Z=6 gap in neutron-rich oxygen isotopes relative to $^{16}$O as a function of the neutron number. The open circles correspond to previous measurements \cite{Mairle16O,Bechtold,Mairle18O}, while the full circles correspond to our work. Error bars corresponding to 1$\sigma$ are shown by a grey shadow region.  Results from the SFO-tls interaction, computed up to the same excitation energy as that measured experimentally, are shown by the red dashed line.}
\end{figure}

The evolution of the $0p$ SO-splitting ($\Delta_{SO}$) is shown in Fig. \ref{espe} for neutron-rich oxygen isotopes.  To avoid any bias in the results, the experimental angular distributions for the proton pick-up of $^{16,18}$O  \cite{Mairle16O,Bechtold,Mairle18O} have been reanalysed using the same optical model parameterizations, overlaps and Wood-Saxon potential parameters as those described in this work\footnote{For $^{16}$O the full strength was measured while for $^{18}$O the strength of 0p$_{1/2}$ and 0p$_{3/2}$ is similar to that obtained for $^{20}$O}. Fig. \ref{espe} shows a clear reduction of the Z=6 shell gap as neutrons are added to the $s_{1/2}d_{5/2}$-orbitals in contrast with  recent results \cite{Ong} suggesting the persistence of a strong Z=6 magic number in light neutron-rich systems. As a matter of fact, the $\Delta_{SO}^{exp}$ changes from 7.09(15) MeV in $^{16}$O to 5.30(14) MeV in $^{20}$O so that the spin-orbit splitting is reduced by $\delta_{SO}=1.79(18)$ MeV when adding more neutrons. 
Results from shell model calculations~\footnote{The theoretical values correspond to a total strength of $> 94\%$} using the SFO-tls interaction are in very good agreement with the quenching of the SO-splitting observed experimentally, as $\Delta_{SO}^{th}$ = 6.92 MeV in $^{16}$O and $\Delta_{SO}^{th}$ = 5.00 MeV in $^{20}$O with $\delta_{SO}=1.92$. The $0p_{1/2}-0p_{3/2}$ splitting can be related to the monopole term of the proton-neutron interaction $\Delta V^{pn}$. 

\begin{equation}
\Delta_{SO}(^{16}O)-\Delta_{SO}(^{20}O)=\Delta V^{pn} = \delta_{SO}
\end{equation}

With $\delta_{SO}$ :


\begin{equation}
\begin{split}
\delta_{SO} = 
&\;\nu_{0d_{5/2}}\big(V^{pn}_{0p_{1/2}0d_{5/2}} - V^{pn}_{0p_{3/2}0d_{5/2}}\big) \\
&+ \nu_{1s_{1/2}}\big(V^{pn}_{0p_{1/2}1s_{1/2}} - V^{pn}_{0p_{3/2}1s_{1/2}}\big)
\end{split}
\end{equation}

Where  $\nu_{0d_{5/2}}$ and $\nu_{1s_{1/2}}$ are the neutron occupancies in the $d$ and $s$ orbitals, respectively. 
The quenching of the spin-orbit splitting appears to be affected by two terms of the monopole interaction that act in opposite directions. While the term multiplied by $\nu_{0d_{5/2}}$ is induced by the tensor force and tends to reduce the  SO-splitting.  The one multiplied by $\nu_{1s_{1/2}}$ corresponds to the action of the two-body LS term that increases the SO-splitting. Therefore, it is important to know how the 4 valence neutrons are distributed across the $s$ and $d$ orbitals. In the context of the experiment, the neutron-removal reaction $^{20}$O(d,t) was measured simultaneously and the deduced spectroscopic factors provided information on the $\ell=0/2$ mixing.  The measured values \cite{Lozano-priv} show that the majority of neutrons are found in the $0d_{5/2}$ orbital, with only 10\% in the $1s_{1/2}$ orbital. Given the good agreement with SFO-tls, one can use the theoretical two-body matrix elements to determine the dominant component. Using spin-tensor decomposition, the difference in the proton 0$p$ SO-splitting is found to arise predominantly from the tensor component with 95\% contribution.

\begin{table}
\centering
\ra{1.2}
\caption{Spin-parity assignments $J^\pi$, $nlj$-orbital, excitation energies $E_{x}$ and spectroscopic factors $C^2S$ for the states observed in $^{20}$O, compared to shell model predictions with the SFO-tls interaction.  }

\vspace{0.15cm}

\begin{tabular}{@{}l l c c c c c@{}}

\hline
\hline
 && \multicolumn{2}{c}{Exp}  & \phantom{a} &\multicolumn{2}{c}{SFO-tls}\\
\hline
$J_i^{\pi}$  &  $n \ell j$  & $E_{x}$ (MeV)& $C^{2}S(0^{+})$  &  &$E_{x}$ (MeV)  &$C^{2}S(0^{+})$  \\
\hline

$1/2_{1}^-$& $0p_{1/2}$  &  0.00(5) &1.66 (11)   & & 0.0     & 1.73  \T \\
$3/2_{1}^-$& $0p_{3/2}$  & 1.21(12)  & 0.49(5)    & & 1.559 & 0.49  \\
$1/2_{2}^-$& $0p_{1/2}$  &  -       & -              & & 3.226 & 0.03 \\
$3/2_{2}^-$& $0p_{3/2}$  & 2.58(9)  & 0.38(5)    & & 3.344 & 0.28  \\
$3/2_{3}^-$& $0p_{3/2}$  & 3.85(15)  & 0.10(2)    & & 4.056 & 0.69  \\
$3/2_{4}^-$& $0p_{3/2}$  & 5.30(3)  & 1.05(7)    & & 4.861 & 0.08  \\
$3/2_{5}^-$& $0p_{3/2}$  & 7.29(9)  & 0.39(4)    & & 5.281 & 0.81  \\
$3/2_{6}^-$& $0p_{3/2}$  &9.39(16)  & 0.28(4)    & & 6.597& 0.43  \\
$3/2_{7}^-$ &  $0p_{3/2}$   & 11.96(26) & 0.18(3)   & &7.826 & 0.09  \\
$3/2_{8}^-$ &   $0p_{3/2}$ & - & -   & &8.243 & 0.06 \\
$1/2_{4}^-$ &  $0p_{1/2}$   & - & -    & &8.656 & 0.04 \\
$3/2_{9}^-$ &   $0p_{3/2}$ & - & -   & &8.885 & 0.49 \\
$3/2_{10}^-$ &   $0p_{3/2}$ & - & -   & &10.626 & 0.17 \\
$3/2_{11}^-$ &   $0p_{3/2}$ & - & -   & &12.093 & 0.10 \B\\

\hline
\hline
\label{tab_sf_19n}
\end{tabular}
\end{table}%

\textit{Conclusion-} 
The single-particle energies of the  $0p_{3/2}$ and  $0p_{1/2}$ orbitals in $^{20}$O have been deduced experimentally. A modification of the SO-splitting of approximately 1.8 MeV has been observed from $^{16}$O to $^{20}$O. This result is at odds with a previous publication claiming a persistent Z=6 shell gap \cite{Ong}. The results are in good agreement with state-of-the art shell model interaction SFO-tls. The reduced 0$p$ SO-splitting is fully attributed to the effect of the tensor force. The ACTAR TPC set-up was crucial for attaining unprecedented resolution for such target thickness and significantly increasing the luminosity, enabling the high-precision measurements presented here.

\textit{Data availability-} 
Experimental data can be found at Ref. \cite{e796_data}.

\textit{Acknowledgments} J.L.F acknowledges financial support from Xunta de Galicia (Spain) grant number ED481A-2020/069. This work was supported by the: Spanish MINECO PID2021-128487NB-I00,  Xunta de Galicia 2021-PG045 and the Maria de Maeztu Unit of Excellence MDM-2016-0692.  The authors acknowledge the support provided by the technical staff of GANIL. This project has received funding from the European Union’s Horizon 2020 research and innovation programme under grant agreement No 654002 (ENSAR2). The authors acknowledge the support of the ENSAR2 Transnational Access programme for providing access to GANIL. J.C.Z. thanks the Fundação de Amparo a Pesquisa do Estado de São Paulo (FAPESP) Grant No. 2018/04965-4. V.G thanks the National Council for Scientific and Technological Development (CNPq) (Grant No. 304961/2017-5).

\end{document}